\documentclass{an}
\usepackage{graphicx}
\usepackage{times}

\begin{document}

\Pagespan{1}{}
\Yearpublication{}
\Yearsubmission{}
\Month{}
\Volume{}
\Issue{}
\DOI{}

\title{The Cygnus X region XXIII. Is 18P87 galactic or extragalactic?}

\author{Otto P. Behre\inst{1}
        \and Heinrich J. Wendker\inst{1,2}
        \and Lloyd A. Higgs\inst{3} 
        \and Thomas L. Landecker\inst{3}}
\institute{Hamburger Sternwarte, Gojenbergsweg 112, D-21029 Hamburg,
           Germany 
         \and It is with deep sadness that we record the death of
Professor H.J. Wendker at the age of 69 years after a long illness
          \and National Research Council, Herzberg Institute of
           Astrophysics, Dominion Radio Astrophysical Observatory, 
           Box 248, Penticton, B.C. V2A 6J9, Canada}
\thanks{Corresponding author: O.P. Behre, e-mail OPBehre@t-online.de}

\received{01 April 2007} \accepted{01 April 2008}
\publonline{} 


\abstract{The radio source 18P87, previously thought to be a point
source, has been serendipitously found to be resolved into a core-jet
geometry in VLA maps. \ion{H}{i} absorption of continuum emission (in
data from the Canadian Galactic Plane Survey) appears in gas with
radial velocities $>+2$ km/s but not in brightly emitting gas at lower
radial velocity. Examination of further archival observations at radio,
infrared and optical wavelengths suggests that the ``obvious''
interpretation as a radio galaxy requires a rather unusual object of
this kind and a highly unusual local line of sight. We argue that
18P87 may be a Galactic object, a local astrophysical jet. If this is
correct it could have arisen from outbursts of a microquasar. }

\maketitle

\newcommand{\pen}{18P87}

\section{Introduction}

Our surveys of the Cygnus X region (Wendker, Higgs, \& Landecker
\cite{cx18}, Paper XVIII) generated several follow-up observations at
higher resolution with the VLA. In one of these we serendipitously
found that the source \pen\ was resolved into a structure that at
first glance resembles a faint radio galaxy and we did not
investigate it more thoroughly. However, while perusing continuum
absorption by \ion{H}{i} in data from the Canadian
Galactic Plane Survey (CGPS - Taylor et al. \cite{art}) we found that
\pen\ shows \ion{H}{i} absorption for only a small part of the local
gas, which is quite unexpected for an extragalactic source. In this
paper we present the available observations and
examine the possibility that the source is Galactic.

\section{Observations}

\subsection{18P87 in CGPS and other survey data}

\begin{figure}
\resizebox{\hsize}{!} 
{\includegraphics{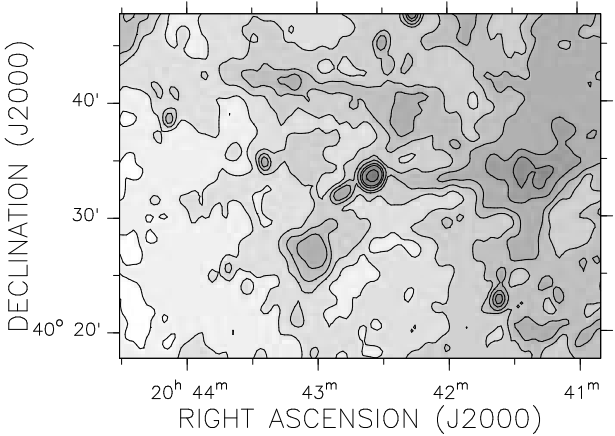}}
\caption{Continuum map of the CGPS area of \pen\ at 21 cm. 
\pen\ is the double source aligned NW-SE at map centre. 
} 
\label{cgps21}
\end{figure}

The large-scale brightness distribution around \pen\ can be seen in a
21-cm continuum map (Fig.~\ref{cgps21}) taken from the CGPS (for which
observation and imaging techniques are described in detail in Taylor
et al. \cite{art}). Observational details relevant to
Fig.~\ref{cgps21} are given in Table~\ref{tab-drao}. \pen , first
catalogued at 408 MHz in Paper XVIII, is the double source near the
centre of the image; it is clearly resolved in Fig.~\ref{cgps21} at a
resolution of $\sim$1\arcmin\ into north-west and south-east components
(\pen NW and \pen SE). The remaining resolved structure in this image
is faint \ion{H}{ii}.

\begin{table}[h]
\caption{\label{tab-drao} Parameters of the DRAO CGPS data.}
\begin{tabular}{ll} 
\hline 
CGPS mosaic MN1 & (release April 2002) \\
Centre frequency & 1.420\,\,GHz\\
Data product & Stokes I\\
Synthesized beam & 71.5\arcsec $\times $ 49.1\arcsec ,\,PA $-$36.2\degr \\
rms noise & 40 mK \\
Line synthesized beam & 85.4\arcsec $\times $ 58.6\arcsec ,\,PA $-$36.0\degr \\
rms noise & 2.5 K \\
Radial velocity coverage & $-$164 to +58 km/s\\
Channel spacing & 0.82446 km/s \\
Velocity resolution & 1.32 km/s\\
\hline
\multicolumn{2}{l}{Data available from www1.cadc.hia.nrc.gc.ca/cgps}
\end{tabular}
\end{table}

The general area, in the south-east of the Cyg-X region, is less
crowded with bright \ion{H}{ii} regions than other parts of the
complex. This eases the derivation of point-source parameters, but the
error budget is still dominated by interpolation of the local
background.  As the resolution of these surveys (and other archival
data that we have used) is around 1\arcmin\ and both components of the
double source appear to be unresolved, we have fitted Gaussians to the
data using the DRAO software routine FLUXFIT.  Equatorial and Galactic
co-ordinates are listed in Table~\ref{pos3}.  In Table~\ref{flux3} we
summarize the flux densities of both components, either taken from
Gaussian fits to the quoted survey data or directly from the
references. The VLA flux densities are integrated values; see Sect.\
\ref{vlaobs}.

The sources are definitely nonthermal. Least squares fitted spectral
indices ($S_{\nu }\propto {\nu ^{\alpha}}$) to all data are $-$1.1 and
$-$0.7 for \pen NW and \pen SE respectively with errors around $\pm
$0.3 dominated by the scatter of the low-frequency values.

\begin{table}[h]
\caption{\label{pos3} Equatorial and Galactic positions of \pen .}
\begin{tabular}{ll}
\hline
Component & Position \\
18P87NW & 20\fh\,42\fm\,34.8\fs\,/+40\degr\,33\arcmin\,43\arcsec ~~(J2000)  \\
        & 80.6928\degr /$-$1.0699\degr  ~~(l,b) \\
18P87SE & 20\fh\,42\fm\,49.4\fs\,/+40\degr\,32\arcmin\,18\arcsec ~~(J2000)  \\
        & 80.7026\degr /$-$1.1207\degr  ~~(l,b) \\
\hline
\end{tabular}
\end{table}
\begin{table}[t]
\caption{\label{flux3} Summary of flux densities of \pen .}
\begin{tabular}{crrcl}
\hline 
Frequency  & \pen NW     & \pen SE           & Obs. & Ref. \\
 $\mathrm{[MHz]}$ & [mJy]     & [mJy]        & Date & \\
\hline
  327      & 234$\pm $10 & 40$\pm $8         & 1991        & 1   \\
  365      & \multicolumn{2}{c}{$<400$}      & 1974-1983   & 2   \\
  408      & \multicolumn{2}{c}{500$\pm $70} & 7-8/1985    & 3   \\
  408      & \multicolumn{2}{c}{423$\pm $36} & 12/95-01/96 & 4   \\
 1400      & 76$\pm $3 & 11$\pm $2           & 28/04/1995  & 5   \\
 1420      & 94$\pm $3 & 17$\pm $3           & 12/95-01/96 & 4   \\
 1465      & 62$\pm $3 & 16$\pm $2           & 17/05/1988  & 4   \\
 4800      & $<20$ & $<20$                   & 05/74       & 6   \\  
 4850      & $<18$ & $<18$                   & 11/86-10/87 & 7   \\  
 4885      & 33$\pm $2 & 6$\pm $1            & 22/08/1988  & 4   \\
\hline
\multicolumn{5}{l}{1 WENSS, Rengelink et al. (\cite{wenss}); 2 Douglas et al. (\cite{texas});} \\
\multicolumn{5}{l}{3 Paper XVIII, Wendker et al. (\cite{cx18}); 4 CGPS, this paper;} \\
\multicolumn{5}{l}{5 NVSS, Condon et al. (\cite{nvss}); 6 Paper XV, Wendker (\cite{cx15});} \\
\multicolumn{5}{l}{7 Gregory et al. (\cite{gb6}) }
\end{tabular}
\end{table}
\subsection{\ion{H}{i} continuum absorption \label{abshi}}

\begin{figure} 
\resizebox{\hsize}{!} 
{\includegraphics{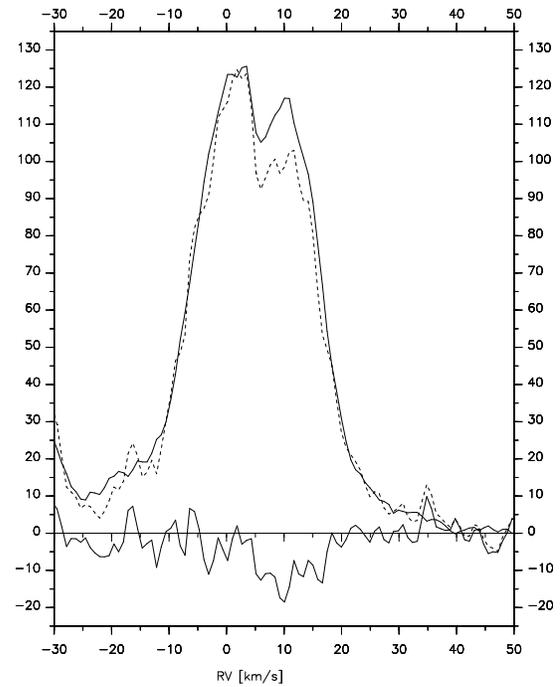}}
\caption{The \ion{H}{i} continuum absorption by local gas is
shown for \pen NW. The top solid line is the off-spectrum, the dashed
line the on-spectrum, and the bottom solid line the difference, on $-$
off. }
\label{hiabs}
\end{figure} 
\begin{figure} 
\resizebox{\hsize}{!}
{\includegraphics{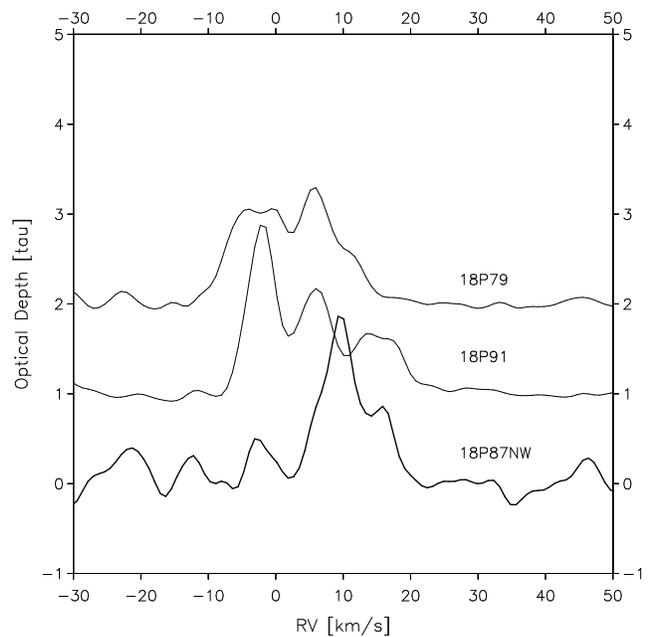}}
\caption{Optical depth spectra for \pen NW, 18P91 (offset
by ${\tau}={1}$), and 18P79 (offset by ${\tau}={2}$) . These spectra have been smoothed by fitting spline functions (see text).}
\label{tau}
\end{figure} 

Even quite weak continuum sources show absorption by \ion{H}{i} in the
CGPS 21-cm line data. Strasser \& Taylor (\cite{absex}) discuss this
for a large sample of {\it{bona fide}} extragalactic sources.
Following the method outlined in Behre et al. (\cite{cx22}, Paper
XXII) we derived the absorption spectrum of \pen\ and several other
(extragalactic) point sources in the neighbourhood. \pen SE is
too weak to produce a believable optical depth signal, but \pen NW is
strongly absorbed in the local gas. In Fig.~\ref{hiabs} we show
on-source and off-source spectra for \pen\ together with the
difference spectrum, on minus off. Significant absorption is evident
at velocities +2 to +18 km/s. In Fig.~\ref{tau} we examine the data as
an optical depth spectrum, together with similar spectra for the
nearby comparison sources 18P79 and 18P91, both about half a degree
distant.  These spectra have been smoothed by fitting a cubic spline
function to each. The rms noise, $\sigma$, in each unsmoothed spectrum
was estimated from channel-to-channel variations, and a spline
function was fitted such that the rms deviation between smoothed and
unsmoothed data was less than $\sigma$.  Values of $\sigma$ are 0.15,
0.059, and 0.19 for \pen NW, 18P79 and 18P91 respectively.

18P79 and 18P91 show significant optical depths, $\tau$, typically 1
to 2, in all local \ion{H}{i} (velocities +25 to $-$15 km/s) and
indications of absorption in the Perseus and outer spiral arms (at
velocities more negative than those shown in Figs.~\ref{hiabs}
and~\ref{tau}). \pen NW shows significant optical depth (up to
${\tau}={1.8}$) but there is an important difference between \pen NW
and the other (extragalactic) sources. For the latter \ion{H}{i}
absorption is seen at {\it all} local velocities, while \pen NW shows
absorption only down to a radial velocity of about +2 km/s, just
including the self-absorption feature which produces the apparent
split of the emission peak into two maxima (Fig.~\ref{hiabs}). There is
a remarkable absence of absorption corresponding to the very strong
peak at $\sim$0~km/s in the emission spectrum (see Fig.~\ref{hiabs}),
and this suggests very strongly that \pen\ lies {\it{within}} the
local gas.

\subsection{High-resolution VLA observations} \label{vlaobs}

In VLA\footnote{The National Radio Astronomy Observatory is a facility
of the National Science Foundation operated under cooperative
agreement by Associated Universities, Inc.}  observations taken for
another purpose we discovered that \pen\ is resolved into a two-lobed
structure. Telescope parameters for 1465 and 4885~MHz are summarized
in Table \ref{tab-vla}. At 4885~MHz low- and high-resolution images were made. As
\pen\ is situated off-centre in both observations, the maps were
corrected for primary-beam attenuation using the standard VLA
prescriptions. In an image made simultaneously at 8415 MHz
the source is too far down the primary beam to allow derivation of
useful data.

\begin{table}[h]
\caption{\label{tab-vla} Parameters of VLA observations.}
\begin{tabular}{ll}
\hline
Field centre 1465 MHz: & \\
$\alpha $(B1950) & 20\fh\,41\fm\,01.0\fs \\
$\delta $(B1950) & 40\degr \,19\arcmin \,00\arcsec  \\
On-source time C+D & 37 min \\
Synthesized map size & 42.58\arcmin\ $\times$ 42.58\arcmin \\
Synthesized beam & 20.9\arcsec $\times $ 19.0\arcsec , PA 61\degr \\
Surface brightness conversion & 1 Jy/beam $\leftrightarrow $ 
1434 K \\
 & \\
Field centre 4885 MHz: & \\
$\alpha $(B1950) & 20\fh\,41\fm\,01.0\fs \\
$\delta $(B1950) & 40\degr \,19\arcmin \,00\arcsec  \\
On-source time C+D & 37.4 min \\
Synthesized map size & 12.77\arcmin\ $\times$ 12.77\arcmin \\
Synthesized beam 1 & 13.2\arcsec $\times $ 10.8\arcsec , PA 64\degr \\
Surface brightness conversion & 1 Jy/beam $\leftrightarrow $ 
359.3 K \\
Synthesized beam 2 & 5.5\arcsec $\times $ 4.4\arcsec , PA 175\degr \\
Surface brightness conversion & 1 Jy/beam $\leftrightarrow $ 
2117 K \\
 & \\
C configuration; 20 and 6 cm & 1988 May 17 \\
D configuration; 20, 6, 4 cm &  1988 August 22 \\
Calibrators: & \\
~~flux density & 3C 286 \\
~~phase & 2050+364 \\
~~polarization & 3C 138 \\
\hline 
\end{tabular}
\end{table} 

\begin{figure}
\resizebox{\hsize}{!}
{\includegraphics{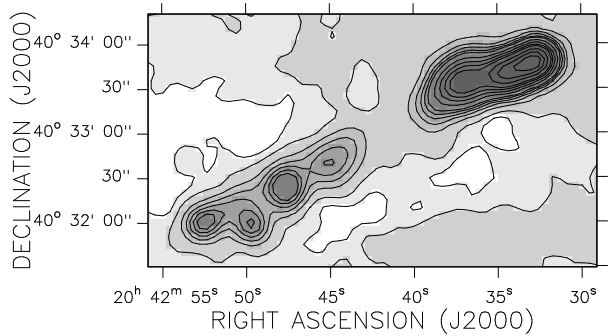}}
\caption{1465 MHz VLA map of \pen, resolution $\sim$20\arcsec\ . The
contours are manually selected to show the salient
features. The contours and the colour transitions correspond to -0.5,
0, 1, 2, 3, 4, 5, 7.5, 10, 12.5, 15, 20, 25, and 30 K.} 
\label{vla20map}
\end{figure}

\begin{figure}
\resizebox{\hsize}{!}
{\includegraphics{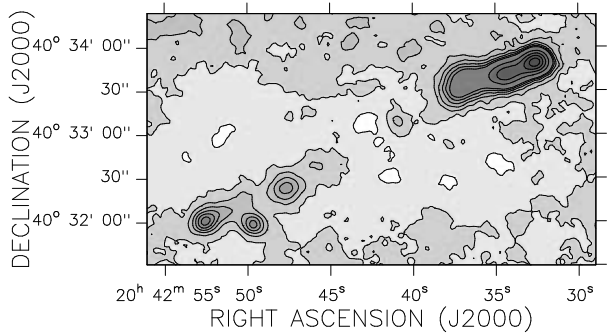}}
\caption{4885 MHz VLA map of \pen, resolution $\sim$13\arcsec\ . The
contours are manually selected to show the salient features. A
central source can be seen. The contours and the colour transitions
correspond to -0.10, 0.0, 0.1, 0.2, 0.3, 0.4, 0.5, 1.0, 1.5, 2.0, and
2.5 K (colour in online edition).
 } 
\label{vla6map}
\end{figure}

VLA maps of \pen\ are shown in Figs.~\ref{vla20map} and
~\ref{vla6map}. At typical resolutions of several arcseconds \pen\ is
resolved into a two-lobed structure about $\sim$5\arcmin\ in overall
length and $\sim$0.5\arcmin\ in width. At both 1465 and 4885~MHz
data from C and D configurations of the VLA were combined, and
{\it{u-v}} coverage allowed complete imaging of objects of this size
without any loss of broad structure.  At first glance the structure
is typical of a radio galaxy with two jets. The images can
be approximated by a series of aligned knots with typical extent
$\sim$10\arcsec\ immersed in a faint elliptical envelope around each
lobe.  The NW lobe contains three such knots and the (longer) SE lobe
four. We will show that the fifth knot in the SE lobe, not aligned
with the dominant source structure, is an unrelated foreground object.
The 1465 and 4885 MHz flux densities listed in Table~\ref{flux3} are
integrated values from Gaussian fits. Flux densities derived from the
low- and high-resolution maps at 4885~MHz agree closely, and the
difference in appearance arises only from resolution effects.  At
4885~MHz a faint knot is situated more or less centrally between the
two lobes; we will refer to it as the central knot.  Its deconvolved
size is 2 to 3\arcsec at 4885~MHz. At 1465~MHz it is not detected,
probably because its spectral index is flatter relative to the lobes,
but also because of the slightly poorer angular resolution.


Spectral indices derived from the integrated flux densities of the
lobes between 1465 and 4885~MHz are quite similar, $-$0.52 and $-$0.81
for the NW and SE lobes respectively, but these values may be
affected by variability (see below). We produced a pixel-by-pixel
spectral index map at the resolution of the 1465~MHz map; while the
errors are large the mean values are consistent with the above
estimates.  However, there is a strong and consistent trend of rising
spectral index from the central knot ($\alpha \sim -1.5$) to the edge
($\alpha \sim -0.5$).  A similarly definite trend appears in spectral
indices derived from fitted Gaussian components; individual values are
noisy but the trend is there.

\subsection{Other observations \label{other}}

Central sources of jets are often variable, and we have search\-ed the
available data (presented in Table~\ref{flux3}) for signs of such
activity. We consider the 6-cm data first. The errors quoted on the
new measurements are from the fitting procedure alone. The older data
do not contribute much, as they were obtained with lower resolution
than our VLA maps. For the bright component (\pen NW) variations at
the 5 to 6$\sigma $ level could have escaped detection.  At this
wavelength we can say only that the source could have brightened
slightly over the time span (1974 to 1988) of the available data.

Examining the 21-cm data, we find that the values for \pen SE are
compatible with no variability, although there is a small probability that
the measured values could have been influenced by activity of the
foreground star (see Sect. \ref{star}). On the other hand, \pen NW
appears to have brightened in the course of 8 years. We are
comparing observations VLA data from 1988 with data from the DRAO Synthesis 
Telescope obtained in 1996 (see Table~\ref{flux3}) . The two datasets are aperture-synth{\-}esis 
observations of an extended source made with slightly different {\it{u-v}} 
plane coverages, but these differences are of no practical consequence 
for observations of 18P87NW whose extent is ${\sim}2'$, and we consider 
the evidence for variability at 21 cm  is strong.
Light time arguments immediately imply that the brightening cannot
have affected the whole extended source but must have arisen from a
small component, either from one of the knots visible in the VLA maps
or from an outburst in the central source.

Synchrotron radiation is intrinsincally highly polarized. Examination
of the CGPS polarization data at 1.4 GHz (Landecker et al. in prep.)
shows features of extent a few arcminutes in the surroundings of \pen\
whose polarized intensity is $\sim$1.3~mJy/beam, but no polarized
feature can be unambiguously associated with \pen .  This translates
to an upper limit of 1.4\% for \pen .

Examination of the POSS and 2MASS surveys reveals no optical
or infrared structures or point sources which could be identified with
the lobes or their substructures; for the one exception see Section
\ref{star}.  The same is true for the near and far infrared bands of
the IRAS and MSX maps.  The central knot has no apparent optical or
near-infrared counterpart.

\subsection{A stellar foreground source \label{star}}

The VLA maps at 1465 and 4885 MHz (Figs.~\ref{vla20map} and
\ref{vla6map}) show one knot which is conspicuously not aligned with
the main axis of the radio lobes.  Its peak position agrees quite well
with the position of a star on the Palomar Observatory Sky Survey as
well as in the 2MASS catalogue (J2000 position in the USNO B1.0 (Monet
et al. \cite{usno}): 20\fh\,42\fm\,49.86\fs\, 
+40\degr\,31\arcmin\,57.9\arcsec ). 
The colours deduced from these catalogues are
compatible with an unreddened late-type foreground star of type around
M0V with a wide range from early K to late M possible (Cox
\cite{cox}). Its distance would be several tens of pc but definitely
below 100 pc. Its detection at radio wavelengths suggests the star
underwent a radio flare at the time of the VLA observations.

\section{Galactic or extragalactic?}

\pen\ definitely is an astrophysical jet, but whether it is Galactic
or extragalactic is not obvious. We compare the two possibilities
based on the review of microquasars by Mirabel \& Rodriguez
(\cite{miro}) and on the description of the prototypical radio galaxy
Cyg A by Carilli \& Barthel (\cite{cyga}). We also draw on the
work of Fernini (\cite{fernini}) and O'Dea et al. (\cite{odea}), who have
examined the properties of Fanaroff-Riley Class II objects (Fanaroff
\& Riley, \cite{frii}).  Radio galaxies are sufficiently abundant
that finding one close to the Galactic plane (latitude about $-$1\degr)
is quite plausible.  On the other hand, Galactic jets are
preferentially situated close to the Galactic plane. Both
possibilities remain open.

The most important argument favouring a Galactic object is provided by
the \ion{H}{i} absorption (Sect.~\ref{abshi} and Figs.~\ref{hiabs} and
\ref{tau}).  \ion{H}{i} continuum absorption of \pen\ is seen only for
radial velocities larger than +2 km/s although both the on and off
signals from local gas at more negative velocities are
stronger. At +2km/s the optical-depth spectrum of \pen NW
(Fig.~\ref{tau}) shows a value of ${\tau}{\approx}{0.3}$, no more than
$2{\sigma}$, while local gas at +10~km/s is producing a $10{\sigma}$
absorption. For the comparison sources 18P79 and 18P91 virtually the
same local gas is producing absorption at the level of 7 to
$30{\sigma}$. If the same gas was in the line of sight to \pen\ there 
is no doubt that the observations would have had the sensitivity to detect 
absorption in it.

The usual interpretation of such behaviour is that the continuum
source must be in front of all gas with a smaller radial
velocity. Hence the source must be Galactic.  In view of the weight of
this argument we tested whether the assumption inherent in deriving
the absorption profile, that the scale of the \ion{H}{i} gas is
substantially larger than the extent of the source, is justified.  We
split the circumference of the ring defining the off profile into four
quadrants. The qualitative appearance of the on~$-$~off profile in
Fig.~\ref{hiabs} was unchanged. Furthermore, along lines of sight to
18P79 and 18P91 (about 20 and 13 pc away at 2 kpc, respectively) we
see strongly absorbing gas in the whole local interstellar medium. If
this missing continuum absorption is a property of the \ion{H}{i} gas
this would be a very strange line of sight although numbers could be
forced to produce the profiles. One would have to adjust the spin
temperature above 600 K in a tunnel along the line of sight to \pen\
for radial velocities below +2 km/s, while all other indications point
to 200 K or less. In view of the abundance of cool \ion{H}{i} in this
area of Cyg X we regard this as highly unlikely and conclude that the
probability that \pen\ is Galactic is high.

In discussing a possible extragalactic origin, we compare \pen\
with FRII radio galaxies, those with prominent jet-powered
lobes. Fernini (\cite{fernini}) and O'Dea et al. (\cite{odea}) have
examined a total sample of some 40 objects whose linear extent ranges
from $\sim$60 to $\sim$600~kpc. If \pen\ fitted into this group of
radio galaxies its distance could lie anywhere from 40 to 400~Mpc.  
The width of the jet of 10\arcsec\ (see Sect. \ref{vlaobs}) would 
correspond to a physical width of 2 to 20~kpc, credible values.

However, considering the morphological evidence, we see that
the detailed geometry of \pen\ is not that of a typical radio galaxy.
Although at first glance it seems to be an edge-brightened
FRII object, it
lacks the typical small-scale `hot spots' in the lobes. In fact there
are no extended lobes and if it is a radio galaxy it has only jets,
with knots of emission lying in the jets (see the morphology of Cyg A
in Carilli \& Barthel 1996). Such a configuration is not observed in
extended radio galaxies to our knowledge, but only on small (pc)
scales in active galactic nuclei. The three pairs of knots in \pen\
are each aligned across the central source and the position angle
shifts by about 5\degr\ going from the inner to the outer pair. This
could easily be three hot spots belonging to a jet of a Galactic
object, marking three consecutive working surfaces in a dense
surrounding interstellar medium with typical sizes of 1.5~pc in
length and about 0.15~pc in intrinsic width. In the extragalactic
interpretation this behaviour is more difficult to accommodate.

Next we consider evidence from the radio spectrum.  Radio galaxies
tend to show a flattening of the spectrum from the central source to
the outside. This is the case with \pen . In radio galaxies the
flattening is caused by old electrons which diffuse backwards from the
working surfaces to the centre and which dominate the diffuse
component in flux. In the case of \pen\ the spectral indices are
dominated by the knots and not by a diffuse component. This is no
problem for a Galactic jet as a slight precession in the jet could
easily make the outer knots the youngest.
The existence of a central source is compatible with both interpretations.  That the central source has no optical or near-infrared counterpart is also plausible. Consider first the interpretation as a radio galaxy. The central core would be surrounded by an accretion disk and a torus, with two jets. The jets appear to lie close to the plane of the sky and the torus, orthogonal to the jet axis, could be in our line of sight to the core. Its dust and molecular gas would then absorb any optical or near infrared emission from the core. If \pen\ is a microquasar we do not expect a strong optical or infrared source at its centre. The temperatures of the accretion disks of microquasars with black holes of mass 1 to 10~M$_{\odot}$ at their cores are much higher than those of AGN. Rees (\cite{rees}) shows that
the disk temperature is ${\rm{T}}\approx{{2}\times{{10^7}~({\rm{M}}/{\rm{M}}_{\odot}})^{-2.5}}$, where M is the black hole mass. Mirabel \& Rodriguez (\cite{miro}) consider this the reason why they could find no optical or infrared counterpart of the central source of the first microquasar, 1E1740.7$-$2942 (Mirabel et al. \cite{mira}; Marti et al. \cite{marti}). We note that \pen\ and 1E1740.7$-$2942 are morphologically similar.

Variability is another consideration. In radio galaxies variability is
usually confined to the central source.  At 21 cm this could possibly
be hidden by resolution effects (although a slight shift in the
position of the NW maximum should have occurred). At 6 cm a total increase
in the flux density of at least 10 mJy occurred between 1987 and
1988. As the flux density of the resolved central source is only
$\sim$1~mJy this increase cannot have occurred there. The discrepancy
can be resolved if the variability lies in one of the (northwestern)
knots; such behaviour would be very unusual for a radio galaxy.

Finally, we consider the observed size of the source.  As a Galactic
source at a distance of 1 kpc, its physical size must be at least
$\sim$1.5~pc (it may be larger, depending on the orientation of the
jets to the line of sight).  This is quite compatible with the size
range of known microquasars: 1E1740.7$-$2942 has a jet $\sim$5~pc in
length, GRS~1758$-$258 has an extent of 6.7~pc, while SS433 spans the
full extent ($\sim$50~pc) of the supernova remnant W50 (Mir\-a\-bel \&
Rodriguez (\cite{miro}) and Rodriguez et al. (\cite{rodr})).

\section{Conclusions}

The hitherto poorly studied radio source \pen\ has been
seren\-di\-pitously shown to have the structure of an astrophysical
jet. We suggest that it may be a
Galactic object. While the distinction on the basis of physical
properties is not totally conclusive, the Galactic interpretation
relies on the absence of
\ion{H}{i} absorption of the continuum emission beyond the middle of
the local gas. If it is indeed a Galactic structure its non-thermal
radio emission classes it as a microquasar.  We suggest that it is
only sporadically active and that the structures we have observed are
the remnants of interaction with a dense interstellar environment.

\acknowledgements 

The Canadian Galactic Plane Survey (CGPS) is a Canadian project with
international partners. The Dominion Radio Astrophysical Observatory
is operated as a national facility by the National Research Council of
Canada. The CGPS is supported by a grant from the Natural Sciences and
Engineering Research Council of Canada. We ackowledge the help of
D. Del Rizzo and R.Kothes at various stages of the study.  This
publication makes use of data products from the Two Micron All Sky
Survey, which is a joint project of the University of Massachusetts
and the Infrared Processing and Analysis Center/California Institute
of Technology, funded by the National Aeronautics and Space
Administration and the National Science Foundation.  This research has
made use of the VizieR catalogue access tool, CDS, Strasbourg, France.
We are grateful to Dr. A.R. Taylor for his thorough refereeing
of this paper.

\end{document}